\documentclass[runningheads]{llncs}
\usepackage{custom}
\begin{document}
\maketitle             
\begin{abstract}
Visual Inspection with Acetic Acid (VIA) remains the most feasible cervical cancer screening test in resource-constrained settings of low- and middle-income countries (LMICs), which are often performed in screening camps or primary/community health centers by nurses instead of the preferred but unavailable expert Gynecologist.
To address the highly subjective nature of the test, various handheld devices integrating cameras or smartphones have been recently explored to capture cervical images during VIA and aid decision-making via telemedicine or AI models.
Most studies proposing AI models retrospectively use a relatively small number of already collected images from specific devices, digital cameras, or smartphones;
the challenges and protocol for quality image acquisition during VIA in resource-constrained camp settings, challenges in getting gold standard, data imbalance, etc. are often overlooked.
We present a novel approach and describe the end-to-end design process to build a robust smartphone-based AI-assisted system that does not require buying a separate integrated device: the proposed protocol for quality image acquisition in resource-constrained settings, dataset collected from $1,430$ women during VIA performed by nurses in screening camps, preprocessing pipeline, and training and evaluation of a deep-learning-based classification model aimed to identify (pre)cancerous lesions.
Our work shows that the readily available smartphones and a suitable protocol can capture the cervix images with the required details for the VIA test well; the deep-learning-based classification model provides promising results to assist nurses in VIA screening; and provides a direction for large-scale data collection and validation in resource-constrained settings.
\end{abstract}
\section{Introduction}
\subsection{Background}
Cervical cancer stands as the fourth leading cause of cancer-related deaths globally.
Its distribution shows great disparity across geographical regions characterized by varying economic conditions, with a staggering 90 \% of such deaths occurring in Low and Middle-Income Countries (LMICs)\cite{gultekin2020world}.
High-income countries reduced the deaths from preventable cervical cancer through effective early precancer detection using pap-smear cytology-based tests and colposcopy tests\cite{zhang2021trends}.
However, these tests have proven difficult to implement and scale in LMICs due to the lack of experts and resources required.

The cytology test is usually triaged with molecular tests to detect human papilloma viruses (HPV) that generally have low specificity but can identify high-risk populations\cite{schiffman2011human}.
Cytology-based tests demand complex infrastructure, high-quality healthcare services, and resources that many low-income countries lack\cite{small2017cervical}.
Similarly, molecular tests are expensive in most resource-constrained settings of the LMICs.
These methods usually require women to make multiple visits to the health center (as sample collection and results do not usually come on the same day), a barrier in rural settings where women must travel several hours or days to arrive at the health center providing such a service.
Visual Inspection with colposcopy requires an expensive device and expert gynecologists, both of which are mostly unavailable in large parts of LMICs.

Visual Inspection with Acetic Acid (VIA) test is an attractive alternative in such resource-constrained settings, as it does not require laboratory equipment and the results come within minutes\cite{who2021guidelinescreening}.
This allows eligible women to be treated immediately by ablating the pre-cancerous tissue in the target cervix region through thermoablation or cryotherapy, leading to an effective single-visit screen-and-treat programs\cite{who2021guidelinescreening}.

Moreover, Self HPV testing, followed by VIA screening has demonstrated improved specificity, mitigating the risk of over-treatment in one-visit screen-and-treat programs\cite{of1999visual}.
The current international guidelines advocate using human papillomavirus (HPV) testing for first-line triage, followed by VIA for low-resource settings\cite{who2021guidelinescreening}.

\subsection{Cervix and Cervical Cancer}
\textbf{The cervix}, a fibromuscular organ, serves as a passageway between the uterine cavity and the vagina, occupying both internal and external positions. 
The junction where the endo and ecto cervix is called the \textbf{Squamocolumnar Junction (SCJ)}. 
SCJ, initially located within the endocervical canal, shifts to the ectocervix during adolescence and pregnancy. The region between the original and shifted SCJ is the \textbf{Transformation zone (TZ)}.
It can be classified into types 1, 2, or 3 based on visibility: Type 1 shows no endocervical components and is fully visible, Type 2 displays some endocervical components but is fully visible, whereas Type 3 includes endocervical components that are not visible.
It is usually the TZ that cervical cancer originates from\cite{prendiville2017colposcopy}.

\subsection{Subjective nature of Visual Inspection with Acetic Acid (VIA)}
VIA involves observation of the cervix for aceto-white changes after application of 3-5\% dilute acetic acid solution. 

The interpretation of aceto-white changes during VIA involves subtle features such as intensity, type of margin, surface characteristics, and the timing of appearance and disappearance of aceto-white. 
Additionally, since the Transformation Zone (TZ) is the origin site of cervical cancer, VIA is inconclusive for the cervix with a completely endocervical TZ (Type 3). Hence, identifying the cervix type is crucial in maintaining the test's specificity.
Owing to these factors, VIA is highly subjective and the diagnostic performance varies significantly based on the expertise of the examiner\cite{jeronimo2007interobserver}. 
In low-income settings where competent experts are often unavailable, nurses and paramedics typically perform the test, posing challenges to its reliability\cite{drummond2017cervical}. 
\\
To enhance the objectivity and reliability of cervical cancer screening, we develop an AI-assisted system that incorporates expert knowledge based on a study in an LMIC. 

\textbf{Our major contributions are as follows}: a. We establish protocols for capturing high-quality images and constructing a robust dataset using standard smartphones during mass screening in community settings, b. We develop labeling and annotation protocols that address challenges related to image quality in such settings, c. Furthermore, we utilize the dataset to develop and evaluate AI-based tools to assist nurses in clinical decision-making and improve VIA decisions' reliability.

\section{Related Work}
Digital cervicography that captures digital images of the cervix following acetic acid application has been explored for quality control and to reduce subjectivity. 
Traditionally, such images were obtained by mounting a digital camera over a colposcope\cite{stafl1981cervicography}.
Still, this method inherits the drawbacks of colposcopy, such as bulkiness, high cost, electricity dependency, and ongoing maintenance requirements. 
Recent advancements in imaging technologies have led to the development of portable devices as alternatives to cameras mounted on colposcopes\cite{lam2015design,cremer2005digital}.
Particularly, the evolution of smartphone technology has created new possibilities for cervical screening in low-resource settings.
Utilizing smartphones as imaging devices during VIA may be the most feasible alternative to enhance the performance of visual inspection with naked eyes, considering the widespread availability and accessibility of smartphones across all demographics\cite{champin2021use,sami2022smartphone}.
 \\
Images acquired through these methods can be utilized beyond just human interpretation and telemedicine, to develop AI systems that assist in decision-making and train healthcare workers, such as nurses and paramedics, who perform VIA in LMICs\cite{sami2022smartphone,desai2022development}. 
In the realm of digital cervicography and colposcopy, there have been efforts to demonstrate how AI models can be leveraged to make decisions regarding VIA outcomes\cite{hu2019observational,zhang2020cervical}, to detect various landmarks in the cervix\cite{guo2020anatomical} as well as to localize lesions\cite{liu2022segmentation}. 
Similarly, AI models trained using images from specialized devices like pocket colposcopes and smartphone endoscopes are beginning to emerge\cite{asiedu2018development,bae2020quantitative}.
Some studies have utilized a portable device called MobileODT\cite{mink2016mobileodt}, which is a smartphone with external attachments and modifications, to develop AI models for tasks such as classifying risk groups based on VIA\cite{xue2020demonstration}, localizing anatomical landmarks\cite{guo2020anatomical}, and ensuring image quality control\cite{guo2020ensemble}.
However, there is sparse literature on works that exclusively use general smartphones to acquire images without any modifications to build AI-assisted systems.
The SEVIA dataset\cite{yeates2016evaluation}, comprising approximately 31,000 images, represents the largest reported dataset of cervix images captured using smartphones.
The data, however, was initially collected for medical training and has only been explored for validating AI systems to identify cervix images\cite{guo2020ensemble}.
\\
Most studies proposing AI models retrospectively use a relatively small number of already collected images from specific devices, digital cameras, or smartphones;
the challenges and protocol for quality image acquisition during VIA in resource-constrained camp settings, challenges in getting gold standard, data imbalance, etc. are often overlooked.
Some studies have reported utilizing smartphones for image or video acquisition to build AI models\cite{kudva2018andriod,ariyani2020pre,vinals2021using} for VIA decision-making. 
Nevertheless, these few existing works have been carried out using small datasets of around $50-300$ images, in controlled settings, or there are challenges in data acquisition such as almost fifty percent of samples are discarded when training AI models with videos due to motion blur and other issues in \cite{vinals2021using} despite capturing the data in clinical facilities.

Our work provides end-to-end protocol design, data collection, model development and prototype evaluation of the system in real-world camp settings where VIA are usually done.

\section{Methodology}
\subsection{Image Acquisition Protocol}
To prepare the dataset, we conducted 32 VIA camps in local health posts of various locations at ***, an LMIC.
We built a custom mobile application that allows users to enter useful information regarding the demographics, and clinical history, allows taking pictures during VIA tests, and the operator's interpretation of VIA outcomes along with images.
After reviewing methods for capturing images using smartphones, digital cameras, and colposcopy from previous studies, we conducted experiments using essential equipment, testing different protocols with dummy cervix models and in initial camps to determine the most suitable method for our settings.
The final image capture protocol established after multiple iterations, along with its underlying rationale, is as follows:
\begin{itemize}
    \item Place the patient in the lithotomy position and insert and adjust the speculum.
    Remove any vaginal discharge using a dry or saline solution-soaked cotton swab.
    \item Initially, we used a stand to capture images as suggested in literature\cite{kudva2018andriod}.
    However, we observed that using a support stand to hold the phone created operational challenges during practical service delivery, particularly due to limitations in the infrastructure of health posts.
    This resulted in a decline in image quality, especially regarding focus and motion artifacts.
    Consequently, the decision was made to adopt a handheld approach as the definitive standard protocol.
    \item To capture the changes in the cervix during VIA, we initially attempted to record videos like Viñals, et al\cite{vinals2021using}.
    However, due to the difficulty in obtaining stable videos using handheld phones, we decided to take multiple images instead of a video; one image before the application of acid and up to six images approximately one minute after the application of acetic acid, with intervals of around 10 seconds between each image.
    \item We initially decided to integrate an internal flashlight, a 2x zoom level, and a combination of manual and tracked focusing (enabled via touch activation), while keeping the camera at a 15 cm distance from the bed.
    However, we discovered that positioning the imaging apparatus as close as possible to the speculum and using zoom functionalities when necessary consistently resulted in better image quality.
    \item Holding the smartphone tilted slightly left or right so that the phone can go closer to the cervix without obstruction from the speculum’s handle was also useful.
    \item Utilizing condoms on the speculum to minimize obstruction from the vaginal walls to the cervix proved to be beneficial in capturing clearer images.
    It was, therefore put into the standard protocol.
\end{itemize}

\subsection{Image Quality Assessment and Annotation Protocol}
\label{sec:image_acq}
We built a web-based quality assessment and annotation platform that is remotely accessible and provides anonymized images uploaded during the camp settings.
Recognizing the significance of quality control procedures, and the fact that not all images and cervix types are eligible for VIA tests, our questionnaire not only captured the VIA decision but also allowed annotators to provide additional details.
These included identifying a. whether the image depicted a cervix, b. flagging any quality issues with a detailed list of potential issues provided as options to click, c. assessing the sufficiency of image quality for other VIA decisions, d. evaluating the visibility of the squamocolumnar junction (SCJ), e. determining decidability based on the visibility of anatomical landmarks, f. noting the orientation, and g. specifying the percentage visibility of the cervix in the image.
\\

\subsection{Pre-processing}
Despite adhering to the image capture protocol, it remained challenging to capture images with only the cervix visible.
Commonly, images contained portions of the speculum or other organs of the woman, including vaginal walls.
We manually labeled 500 images and trained a Faster R-CNN\cite{ren2015faster} architecture-based cervix detection model.
This was then used to automatically crop the input image, restricting the regions primarily to the cervix, and sending it to the next step.

\subsection{ResNet-18-based Classification Model}
A custom model architecture using ResNet-18 as the backbone was built, which takes (as shown in Figure below) a pair of pre-VIA and post-VIA images as input from the same patient.
\begin{figure}
\label{fig:architecture}
    \centering
    \includegraphics[width=0.82\linewidth]{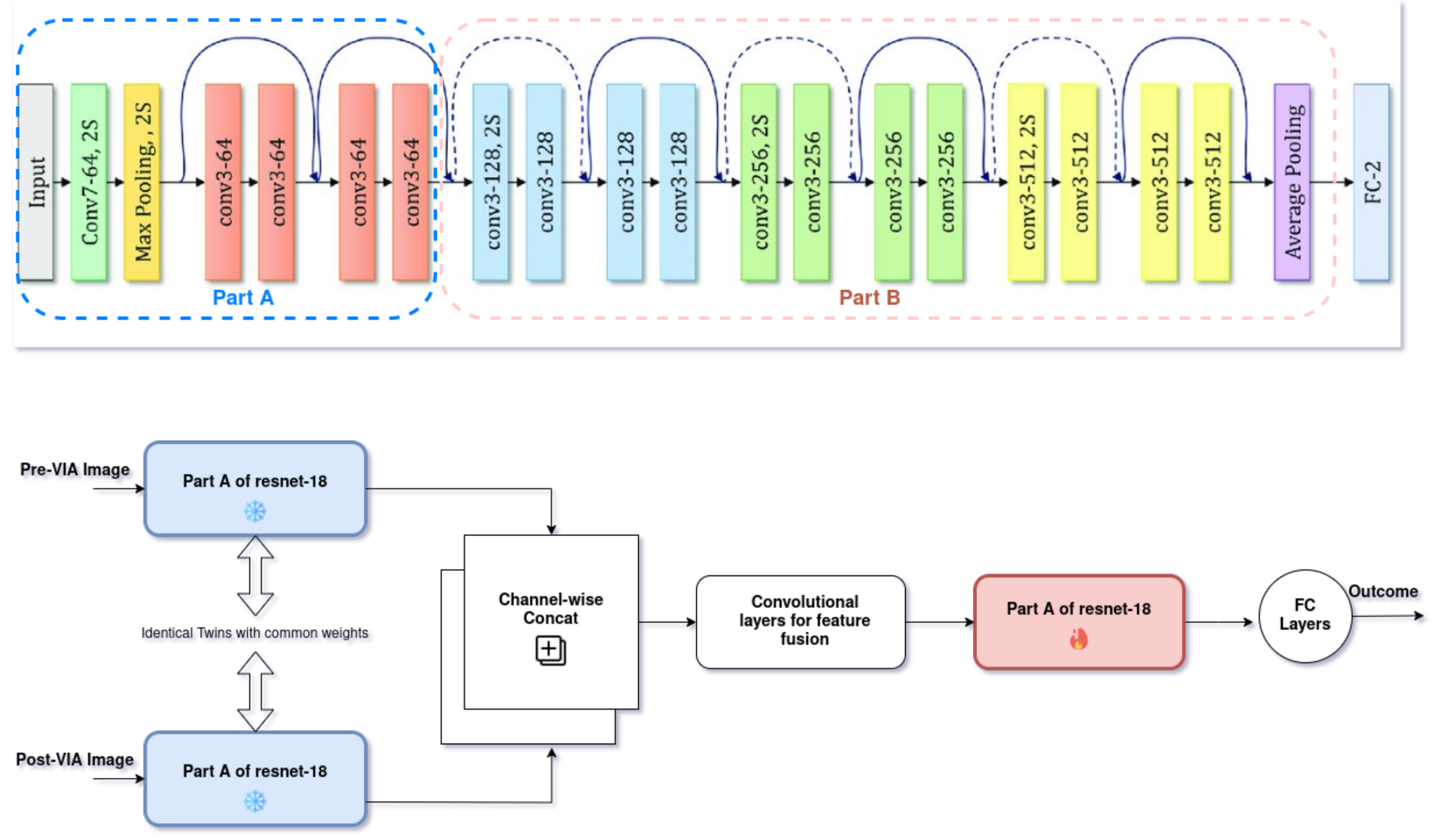}
    \caption{Architecture of the model used(lower block) based on ResNet-18(upper block) 
}
    \label{fig:enter-label}
\end{figure}
\\
The architecture diagram has been presented in Figure\ref{fig:architecture}\footnote{ResNet-18 diagram from: \url{https://www.researchgate.net/figure/Architecture-of-the-ResNet-18-model-used-in-this-study_fig3_354432343}}.
The model takes in $2$ images as input, $1$ images taken before the application of acetic acid, and the other after the application. Both images pass through Part A of ResNet-18, initialized with ImageNet weights, and kept frozen.
The feature maps from Part A of both images, each containing n channels, are concatenated channel-wise, resulting in a feature map with 2n channels.
Additional convolutional layers then reduce the number of channels back to n. This feature map is subsequently passed through Part B of ResNet-18, followed by fully connected layers and a final sigmoid layer to predict the VIA outcome as positive or negative.

\subsection{Implementation Details}
\subsubsection{Quality Control and Dataset Creation}
We collected 9844 images from 1430 women with around 6.8 images from each (one pre-VIA, at most 6 post-VIA), and were annotated by three experts (uniformly distributed) on the participant level.

Based on our questionnaire, experts initially classified post-acetic acid images into four classes: VIA Positive, Negative, Suspicious Cancer, and Not Decidable (based on anatomical visibility), provided they did not flag it as insufficient based on image quality.
We discarded the $1406$ images flagged as of insufficient quality and $30$ of undecidable anatomical visibility as a quality control step.
We combined "Suspicious Cancer" and "VIA Positive" into the "Positive" class.
Finally, 8752 images were used to build model building and internal validation, where the train-test split was done at the patient's level, maintaining the proportion of positive and negative cases. The details of the split are given in Table\ref{tab:data_splits}

\begin{table}[]
\centering
    \caption{Overall Distribution of the Data into train and test sets}
    \label{tab:data_splits}
    
    \begin{tabular}{|l|ll|l|ll|}
    \hline
    \multirow{2}{*}{} & \multicolumn{2}{c|}{\textbf{Women}} & \multicolumn{1}{c|}{\multirow{2}{*}{\textbf{Pre-VIA Images}}} & \multicolumn{2}{c|}{\textbf{Post-VIA Images}} \\ \cline{2-3} \cline{5-6} 
     & \multicolumn{1}{l|}{\textbf{Negative}} & \textbf{Positive} & \multicolumn{1}{c|}{} & \multicolumn{1}{l|}{\textbf{Negative}} & \textbf{Positive} \\ \hline
    Train Set & \multicolumn{1}{l|}{935} & 222 & 1157 & \multicolumn{1}{l|}{5417} & 1203 \\ \hline
    Test Set & \multicolumn{1}{l|}{113} & 33 & 146 & \multicolumn{1}{l|}{657} & 172 \\ \hline
    Total & \multicolumn{1}{l|}{1048} & 255 & 1303 & \multicolumn{1}{l|}{6074} & 1375 \\ \hline
    \end{tabular}

\end{table}
\subsubsection{Experimental Setup}
We utilized a Weighted Binary Cross Entropy Loss for training, assigning a weight of 4 to the positive class to address the data imbalance. 
During training, we applied random online augmentations (flipping, rotation, change of contrast), with a probability of $0.5$ for spatial and $0.2$ for non-spatial augmentations.
15\% of the data was internally split for validation during training.
\\
The model was trained with a batch size of 32. Adam as an optimizer with an initial learning rate of $1e-4$, scheduled with ReduceLROnPlateau, reduced the learning rate by a factor of 0.05 when no improvement in validation loss was obtained for 10 epochs until it reached $1e-6$.
We used early stopping with a patience of 20 epochs to stop the training when no further improvement in validation loss was obtained for $20$ epochs.

\section{Results}

\begin{figure}
    \includegraphics[width=0.35\linewidth]{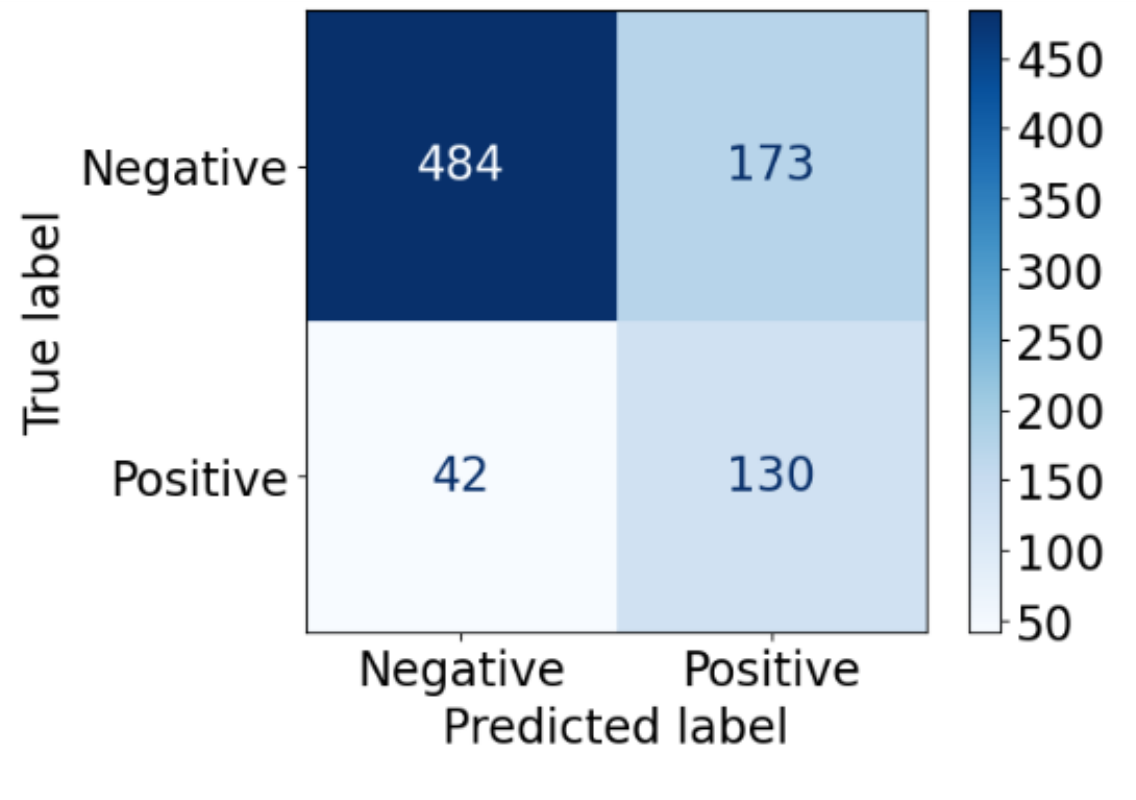}
    \includegraphics[width=0.65\linewidth]{images/mis.pdf}
    \label{fig:outputs}
    \caption{Results: Confusion Matrix on the test set(Left), Visualization of sample images with false predictions(Right); in the Left of each pair is a pre-VIA image and in the right is a post-VIA image.
}
    \label{fig:model_mistakes}
\end{figure}

The model achieved a balanced accuracy of 74.62\%, a sensitivity of 75.58\%, and a specificity of 73.67\% when benchmarked against expert evaluations.
The output confusion matrix is presented in Figure~\ref{fig:outputs}.
78 cases were assigned among the three experts to assess inter-rater reliability. The resulting Krippendorff's alpha score was calculated to be 0.49. 
This score indicates that "the raters reached agreement on 49\% of the samples that were anticipated to have disagreement by chance," demonstrating a relatively low level of agreement. 

Very few patients underwent biopsy from our camp collection, who had suspicious cancer.
Evaluating against biopsy results in this small number of 25 cases (included in the test set of Table\ref{tab:data_splits}), assessing Low-grade squamous intraepithelial lesion (LSIL)+ cases in biopsy as positive, our model exhibited a reasonable sensitivity of $59$\%, but its specificity was relatively low at $25$\%.

\section{Discussion and Conclusion}
Our work proposes an AI solution relevant to LMIC, by taking into consideration the nuances and difficulties in data acquisition.
We propose a protocol allowing any smartphones to be used in resource-constrained VIA camps run by nurses, design and implement an easy-to-use mobile app to collect data with quality control mechanisms, provide annotation, and evaluate VIA decision by a deep-learning-based classification model.
Although approximately 1500 participants is a relatively low number with data imbalance on positive-negative pairs, our dataset does not select only positive samples but rather reflects the proportion of positive-negative samples one would find in realistic settings.
Despite this imbalance, high inter-subject variability, and inherent challenges, the AI model shows the potential to learn features important to classify VIA outcomes in the held-out test set accurately. 
Notably, considering the low agreement between the raters, the labels derived from experts' interpretations may exhibit significant inconsistency and noise, potentially affecting the model's performance.
Low agreement between the raters also reflects the challenges due to the subjectivity of VIA.
Therefore, enhancing label consistency and undertaking further research to address the challenges posed by this dataset could significantly contribute to strengthening the model's reliability and consistency. 
\\
Moreover, our protocol and system allow expanding the dataset in the future to get a much larger number of participants and make a more rigorous validation including prospective trials.

The model's outcome concerning biopsy was also in line with our expectations, given that the model's performance is based on expert interpretation rather than using biopsy as the definitive gold standard and considering the challenges in the dataset mentioned above. 
Additionally, all the cases within this subset originated from the high-risk group, who were initially identified as positive and were subsequently referred for PAP smear/HPV tests followed by Biopsy.
Around 60\% followed up for the PAP smear/HPV tests, and biopsies were conducted for those cases with complicated outcomes in these tests.
Biopsies were not feasible for all VIA-positive cases due to variations in cervix anatomy and adherence to clinical protocols.
\\
Visualizing failure cases, including specific samples shown in Figure\ref{fig:outputs}, underscores the numerous real-world challenges encountered during data acquisition.
These challenges include images tainted with blood or mucus that appear like acetowhite, variations in exposure or brightness levels, instances of specular reflection, etc.
In addition to collecting larger datasets, future work will integrate other ML tasks such as classifying cervix types eligible for VIA outcomes, providing landmarks in real-time to assist nurses in decision-making, and evaluating how these different elements improve nurses' performance. 

%
%
%
\bibliographystyle{splncs04}
\bibliography{references}

\end{document}